\documentclass[fleqn,10pt]{wlscirep}
\usepackage[utf8]{inputenc}
\usepackage[T1]{fontenc}

\usepackage{url,hyperref,lineno,microtype,subcaption}
\usepackage[onehalfspacing]{setspace}
\usepackage{multirow}
\usepackage{amssymb, amsmath}
\usepackage{float}
\renewcommand{\arraystretch}{1.5}

\usepackage{xspace}
\usepackage{xcolor}
\usepackage{comment}
\usepackage{booktabs}

\newcommand{\trans}{\ensuremath{t}\xspace}

\newcommand{\addr}{\ensuremath{a}\xspace}
\newcommand{\addri}[1]{\ensuremath{\addr_{#1}}\xspace}

\newcommand{\ent}{\ensuremath{e}\xspace}
\newcommand{\enti}[1]{\ensuremath{\ent_{#1}}\xspace}

\newcommand{\strans}{\ensuremath{\mathcal{T}}\xspace}
\newcommand{\stransi}[1]{\ensuremath{\strans_{#1}}\xspace}

\newcommand{\sent}{\ensuremath{\mathcal{E}}\xspace}

\newcommand{\saddr}{\ensuremath{\mathcal{A}}\xspace}
\newcommand{\saddri}[1]{\ensuremath{\saddr_{#1}}\xspace}

\newcommand{\sgt}{\ensuremath{\saddr^\star}\xspace}

\newcommand{\ssample}{\ensuremath{\saddr^S}\xspace}
\newcommand{\ssamplei}[1]{\ensuremath{\ssample_{#1}}\xspace}

\newcommand{\ssamplegti}[1]{\ensuremath{\sgt_{#1}}\xspace}
\newcommand{\sentgti}[1]{\ensuremath{\mathcal{E}^\star_{#1}}\xspace}
\newcommand{\ssampleigti}[1]{\ensuremath{\saddr^{(i)}_{#1}}\xspace}
\newcommand{\sentigti}[1]{\ensuremath{\mathcal{E}^{(i)}_{#1}}\xspace}

\newcommand{\ssampleinii}[1]{\ensuremath{\saddr^{I}_{#1}}\xspace}

\newcommand{\stransam}{\ensuremath{\strans^{S}}\xspace}
\newcommand{\stransami}[1]{\ensuremath{\stransam_{#1}\xspace}}

\newcommand{\stransamoc}[2]{\ensuremath{\stransami{[#1,#2]}}\xspace}
\newcommand{\stransamocs}[4]{\ensuremath{\stransami{\intoc{#1}{#2}{#3}{#4}}}\xspace}
\newcommand{\stransampar}[2]{\ensuremath{\stransami{\into{#1}{#2}}}\xspace}
\newcommand{\stransamcum}[2]{\ensuremath{\stransamocs{11}{2}{#1}{#2}}\xspace}

\newcommand{\saddrini}[1]{\ensuremath{\mtrain{#1}}\xspace}

\newcommand{\saddrouti}[1]{\ensuremath{\mtraout{#1}}\xspace}

\newcommand{\sclus}{\ensuremath{\mathcal{C}}\xspace}
\newcommand{\sclusi}[1]{\ensuremath{\sclus_{#1}}\xspace}
\newcommand{\sclusiti}[1]{\ensuremath{\sclus_{[o,c]}^{(#1)}}\xspace}

\newcommand{\scls}{\ensuremath{\mathcal{K}}\xspace}

\newcommand{\sclsiti}[1]{\ensuremath{\scls_{[o,c]}^{(#1)}}\xspace}

\newcommand{\gcornet}{\ensuremath{\mathcal{G}}\xspace}
\newcommand{\gcorneti}[1]{\ensuremath{\gcornet_{#1}}\xspace}

\newcommand{\gcornetedgi}[1]{\ensuremath{\mathcal{L}_{#1}}\xspace}

\newcommand{\maddentgtn}{\ensuremath{\mathbf{e}^\star}\xspace}
\newcommand{\maddent}[1]{\ensuremath{\mathbf{e}(#1)}\xspace}
\newcommand{\maddentgt}[1]{\ensuremath{\mathbf{e}^\star(#1)}\xspace}

\newcommand{\mwghn}{\ensuremath{\mathbf{w}}\xspace}
\newcommand{\mwgh}[1]{\ensuremath{\mathbf{w}(#1)}\xspace}

\newcommand{\mtrainn}{\ensuremath{\mathbf{i}}\xspace}
\newcommand{\mtrain}[1]{\ensuremath{\mtrainn(#1)}\xspace}

\newcommand{\mtraoutn}{\ensuremath{\mathbf{o}}\xspace}
\newcommand{\mtraout}[1]{\ensuremath{\mtraoutn(#1)}\xspace}

\newcommand{\tin}{\ensuremath{T}\xspace}
\newcommand{\tini}[1]{\ensuremath{\tin_{1}}\xspace}

\newcommand{\into}[2]{[#1\text{s}#2]}
\newcommand{\intoc}[4]{[#1\text{s}#2,#3\text{s}#4]}

\newcommand{\msampleinii}{p}

\newcommand{\vE}{\ensuremath{E}\xspace}
\newcommand{\vEi}{\ensuremath{E_i}\xspace}
\newcommand{\vC}{\ensuremath{C}\xspace}
\newcommand{\vCj}{\ensuremath{C_j}\xspace}

\title{The Complex Community Structure of the Bitcoin Address Correspondence Network}

\author[1]{Jan Alexander Fischer}
\author[1]{Andres Palechor}
\author[2,3,*]{Daniele Dell'Aglio}
\author[3,4]{Abraham Bernstein}
\author[4,5]{ Claudio J. Tessone}
\affil[1]{Universit\"at Z\"urich,  R\"amistrasse 71, CH-8006 Z\"urich, Switzerland}
\affil[2]{Department of Computer Science, Aalborg University, Selma Lagerløfs Vej 300, DK-9220 Aalborg, Denmark}
\affil[3]{Department of Informatics, Universit\"at Z\"urich, Binzm\"uhlestrasse 15, CH-8050 Z\"urich, Switzerland}
\affil[4]{UZH Blockchain Center, Universit\"at Z\"urich, Andreasstrasse 15, CH-8050 Z\"urich, Switzerland}
\affil[5]{URPP Social Networks, Universit\"at Z\"urich, Andreasstrasse 15, CH-8050 Z\"urich, Switzerland}
\affil[*]{dade@cs.aau.dk}

\begin{abstract}

Bitcoin is built on a \textit{blockchain}, an immutable decentralised ledger that allows entities (users) to exchange Bitcoins in a pseudonymous manner. Bitcoins are associated with alpha-numeric \textit{addresses} and are transferred via \textit{transactions}.
Each transaction is composed of a set of input addresses (associated with unspent outputs received from previous transactions) and a set of output addresses (to which Bitcoins are transferred).
Despite Bitoin was designed with anonymity in mind, different heuristic approaches exist to detect which addresses in a specific transaction belong to the same entity. 
By applying these heuristics, we build an \emph{Address Correspondence Network}: in this representation, 
addresses are nodes are connected with edges if at least one heuristic detects them as belonging to the same entity. 
In this paper, we analyse for the first time the Address Correspondence Network and show it is characterised by a complex topology, signalled by a broad, skewed degree distribution and a power-law component size distribution. 
Using a large-scale dataset of addresses for which the controlling entities are known, we show that a combination of external data coupled with standard community detection algorithms can reliably identify entities.
The complex nature of the Address Correspondence Network reveals that usage patterns of individual entities create statistical regularities; and that these regularities can be leveraged to more accurately identify entities and gain a deeper understanding of the Bitcoin economy as a whole. 
\end{abstract}
\begin{document}

\maketitle
\section{Introduction}

Cryptocurrencies are rapidly growing in interest, becoming a popular mechanism to perform pseudonymous exchanges between users (\emph{entities}). They also allow payments in a decentralised manner without needing a trusted third party. The first and most popular cryptocurrency is Bitcoin, which uses an immutable and publicly available ledger to facilitate transactions between entities. 
Moreover, given its pseudo-anonymity, Bitcoin has also been used to perform activities in illegal markets. For example, \cite{foley_sex_2019} estimate that one-quarter of entities in the Bitcoin network are associated with illegal activity. Consequently, several governing challenges have arisen, and law enforcement agents are particularly interested in techniques that allow tracing the origin of funds. Specifically, in Bitcoin, given the ledger's public nature, tracing the funds can be achieved by inspecting the history of transactions in the system. However, identifying the entities is a complex task because they can use different pseudonyms (\emph{addresses}) in the system. By the Bitcoin protocol, it is impossible to completely de-anonymise the entities; however, not all entities prioritise anonymity \cite{gaihre_bitcoin_2018}, and it is possible to find recoverable traces of their activity in the transaction history.

The structure of the transactions allows, in some cases, tracing back address pseudonyms that potentially belong to the same entity. For example, \cite{meiklejohn_fistful_2016} apply heuristics and then cluster together pseudonyms based on evidence of shared spending authority. In this paper, we study the application of several heuristics that leads to creating a sequence of \emph{Address Correspondence Networks}. Each of these networks includes weighted links between addresses that potentially belong to the same entity, thus approaching entity identification from a network science perspective. Even though other approaches use networks to model some parts of the Bitcoin economic dynamics (e.g. \cite{kondor_rich_2014, javarone_bitcoin_2018, bovet_evolving_2019}), to the best of our knowledge, network science approaches have not addressed the problem of analysing the Address Correspondence Network to date. In this study, we show that the Address Correspondence Networks have a strong community structure and general-purpose clustering approaches are suitable for analysing them. Furthermore, our experiments suggest that having a set of identified entities generates large gains in cluster quality---however, this gain quickly declines, and a small number of known entities is enough to produce significant increase in the quality of the detection. 

The rest of this paper is organised as follows: section \ref{s:bkg} explains the basics of the Bitcoin blockchain, heuristics, entity identification and related work.
Section \ref{s:method} presents our methods for constructing Address Correspondence Networks, the clustering technique and its quality metrics. In section \ref{s:result}, we discuss our findings, and finally, in section \ref{s:conclusion}, we discuss conclusions and future work.

\section{Background and related work} \label{s:bkg}
This section introduces the main concepts related to Bitcoin.
Next, it discusses the the task of identifying addresses controlled by the same entity, followed by a reviews of the main studies in the area.
\subsection{The Bitcoin blockchain}
Bitcoin was introduced in \cite{nakamoto_bitcoin_2008} as a decentralised payment network and digital currency which would be independent of central bank authorities. 
It is built on a \emph{blockchain}, an immutable decentralised ledger that allows users, i.e. \emph{entities}, to exchange the units of account (\emph{Bitcoins}) in a pseudonymous manner. Entities transacting in the Bitcoin network control \emph{addresses}---unique identifiers which have the right to transfer specific amounts of Bitcoins. 

There are different types of addresses, which determine how the associated Bitcoins are accessed. For example, to spend Bitcoins associated with an address of type Pay to Public Key Hash (P2PKH), the entity needs to present a valid signature based on their private key, and a public key that hashes to the P2PKH value. Another example is the Pay to Script Hash (P2SH) address type: it defines a script for custom validation, which may include several signatures, passwords and other user-defined requirements. We denote with \addr an address and with \saddr the set $\{\addri{1}, \dots, \addri{n}\}$ of addresses appearing in the Bitcoin blockchain. Furthermore, we denote an entity as \ent, with $\mathcal{E}$ representing the set $\{\enti{1}, \dots, \enti{k}\}$ of entities that own Bitcoin addresses.   

To spend or receive Bitcoins, entities create \emph{transactions}. 
A transaction \trans is composed of a set of input addresses, a set of output addresses, and information specifying the amount of Bitcoins to be allocated to each output address. 
Formally, let \strans be the set of transactions stored in the Bitcoin blockchain, and $\mathcal{P}(\saddr)$ be the power set of \saddr.
We model with $\mtrainn : \strans\rightarrow\mathcal{P}(\saddr)$ and $\mtraoutn:\strans\rightarrow\mathcal{P}(\saddr)$ the mappings between a transaction and its input and output address sets.
The sum of Bitcoins associated with the input addresses equals the sum of Bitcoins associated with the output addresses plus transaction fees. 
Therefore, if an entity wishes to spend only a partial amount of Bitcoins associated with the input addresses, the remainder is typically sent to an existing or newly created \textit{change address} controlled by the initiating entity.
Transaction outputs that have not yet been used as inputs to other transactions are referred to as UTXOs (unspent transaction outputs).

The transaction history is replicated on multiple nodes in the Bitcoin network. Entities broadcast new transactions to other nodes in the network. As part of Bitcoin's decentralised consensus protocol, specialised miner nodes are incentivised to solve proof-of-work puzzles that validate new transactions and group them into blocks. Blocks are sequentially appended to the blockchain; the number of blocks preceding a particular block is known as its \emph{block height}. 
Furthermore, entities may specify a transaction's \emph{locktime}. This is the minimum block height the blockchain must reach before miners should consider validating the transaction, i.e. a transaction with locktime $j$ is added to block $j+1$ or later.

A peculiar property of the Bitcoin network is the \emph{pseudonymity}: entities conceal their identity through the use of nameless addresses (pseudonyms), linking an address to a real-world entity exposes their entire activity on the Bitcoin network, since the transaction history is publicly available. Entities are therefore advised to generate a new address for every transaction, so that each address is used once as a transaction output and once as a transaction input.

\subsection{Address clustering}\label{s:bkg:heu}
The objective of \emph{address clustering} is to find sets of addresses $\saddri{i} \subseteq \mathcal{A}$ that are controlled by the same entity $\enti{i}$. Formally, the objective is to find a map $\mathbf{e}:\saddr\rightarrow\sent$ such  that  $\saddri{i} = \{ \addri{j} | \maddent{\addri{j}} = \enti{i}\}$. 
There exist multiple heuristics for identifying address pairs controlled by the same entity. We consider seven heuristics implemented by \cite{kalodner_blocksci_2017}, the majority of which seek to identify change addresses in the outputs of a transaction (linking these with the transaction inputs).
\begin{enumerate}
    \item\emph{Multi-input.} All input addresses of a transaction are assumed to be controlled by the same entity.
    \item\emph{Change address type.} If all input addresses of a transaction are of one address type (e.g. P2PKH or P2SH), the potential change addresses are of the same type. 
    \item\emph{Change address behavior.} Since entities are advised to generate a new address for receiving change, an output address receiving Bitcoins for the first time may be a change address.
    \item\emph{Change locktime.} If a transaction's locktime is specified, outputs spent in different transactions on the same block as the specified locktime may be change addresses. Intuitively, this is because the entity initiating the transaction also knows its locktime.
    \item\emph{Optimal change.} If an output is smaller than any of the transaction inputs, it is likely a change address. 
    \item\emph{Peeling chain.} In a peeling chain, a single address with a relatively large amount of Bitcoins begins by transferring a small amount of Bitcoins to an output address, with the rest being allocated to a one-time change address. This process repeats several times until the larger amount is reduced, meaning that addresses continuing the chain are potential change addresses (\cite{meiklejohn_fistful_2016}).
    \item\emph{Power of 10.} This heuristic assumes that the sum of deliberately transferred Bitcoins in a transaction is a power of 10. If such an output is present, the other outputs may be change addresses.
\end{enumerate}



\subsection{Related work} \label{s:rel_work}
Address clustering in Bitcoin has been the subject of numerous studies. 
Initial studies focused on the \emph{multi-input} heuristic.
For example, \cite{nick_data-driven_2015} identify more than 69\% vulnerable addresses using only this heuristic.  
Also \cite{harrigan_unreasonable_2016} consider the multi-input heuristic and attribute its effectiveness to frequent address reuse, as well as the presence of large address clusters having high centrality measures with respect to transactions between clusters. Furthermore, they suggest that incremental cluster growth and the avoidable merging of large clusters makes the multi-input heuristic suitable for real-time analysis. 
\cite{fleder_bitcoin_2015} construct directed transaction graphs for periods of 24 hours and 7 months. In such graphs, the nodes are addresses and each edge represents a transaction from an input address to an output address. They obtain address entity labels by scraping public forums and social networks. By applying the multi-input heuristic, they identify transactions where labelled addresses have interacted with a large number of known entities such as SatoshiDICE and Wikileaks.

\cite{meiklejohn_fistful_2016} combines the multi-input heuristic with a second one, similar to the \emph{change address behavior} heuristic. 
They identify major entities and interactions between them, and note that the change address heuristic tends to collapse address groups into large super-clusters. 
\cite{zhang_heuristic-based_2020} consider another variation of the change address behavior heuristic, and show that it improves clustering quality when address reduction is used as a performance measure. 
In this study, we focus on the heuristics introduced in Section \ref{s:bkg:heu} by \cite{kalodner_blocksci_2017}.

\cite{patel_deanonymizing_2018} proposes novel approaches to Bitcoin address clustering. 
He considers clustering an undirected, weighted heuristic graph, where the nodes are addresses, and each edge indicates the presence of at least one of eight heuristics (a superset of those introduced in section \ref{s:bkg:heu}) linking those addresses to the same entity.
Each heuristic is assigned a positive weight, such that their sum is equal to one. 
The edge weight is the sum of the heuristic weights for which the corresponding heuristic is present between two addresses. 
The author applies a variety of generic graph clustering algorithms (e.g. $k$-means, spectral, DBSCAN) as well as graph sparsification and coarsening techniques to the constructed heuristic graph. 
In this study, we propose the address correspondence network, which is similar to the network built by \cite{patel_deanonymizing_2018} 
However, in our correspondence network, an edge between two addresses represents the number of times the heuristics identify the pair as controlled by the same entity. We use a label propagation algorithm to build the clusters, using ground truth information to drive the algorithm. 

There exist other approaches and extensions to address clustering. 
\cite{ermilov_automatic_2017} show that higher cluster homogeneity can be achieved when transaction data is augmented with off-chain information from the internet.
\cite{biryukov_deanonymization_2019} propose incorporating lower-level network information to enhance deanonymisation. 
Furthermore, \cite{harlev_breaking_2018} extend address clustering by using supervised machine learning to predict the type of entity controlling addresses in an unlabeled cluster. 
In our study, in addition to using a ground truth to guide the clustering construction, we introduce a temporal component in the analysis. We build address correspondence networks for various time intervals. In this way, we can analyze the evolution of the network over time.

\section{Methodology} \label{s:method}

We expand upon the work of \cite{patel_deanonymizing_2018} by performing address clustering on so-called \emph{Address Correspondence Networks}, denoted $\gcorneti{[o,c]}$, where $[o,c]$ is a time interval. Nodes are Bitcoin addresses that are involved in transactions between a time instant $o$ and a time instant $c$. 
$\gcorneti{[o,c]}$ contains an undirected link $(\addri{i},\addri{j})$ between two addresses when at least one of the heuristics introduced in section \ref{s:bkg:heu} detects \addri{i} and \addri{j} as belonging to the same entity.
We posit that the topology of $\gcorneti{[o,c]}$ encodes further insights on the identity of the entities and, ultimately, on the \maddent{\addri{j}} map. 

For some addresses \addri{j}, the controlling entity is known. 
Using the block explorer tool provided by \cite{janda_walletexplorercom_2017}, we obtain entity labels for 28 million addresses involved in transactions before 2017. We refer to this data set as the \emph{ground truth}.
The mapping information contained in the ground truth is denoted with \maddentgtn, such that $\sgt = \{ \addri{j}  | \exists \maddentgt{\addri{j}}\} \subseteq \mathcal{A}$ is the set of addresses for which the entity label is known. 
We use the ground truth to (1) sample from \strans and (2) to evaluate the quality of address clustering methods. 


The remainder of this section is organised as follows. Section \ref{s:method:samp} describes the method for sampling from \strans. This sample is divided further into cumulative and partial subsets, which are described in section \ref{s:method:subsets}. Section \ref{s:method:corresp} details the construction of the Address Correspondence Networks. We explain our approach to clustering these networks in section \ref{s:method:cluster}, while the metrics used to evaluate clustering quality are introduced in section \ref{s:method:quality}.



\subsection{Transaction sampling} \label{s:method:samp}


\begin{figure}[b!]
    \centering
    \includegraphics[width=0.75\textwidth]{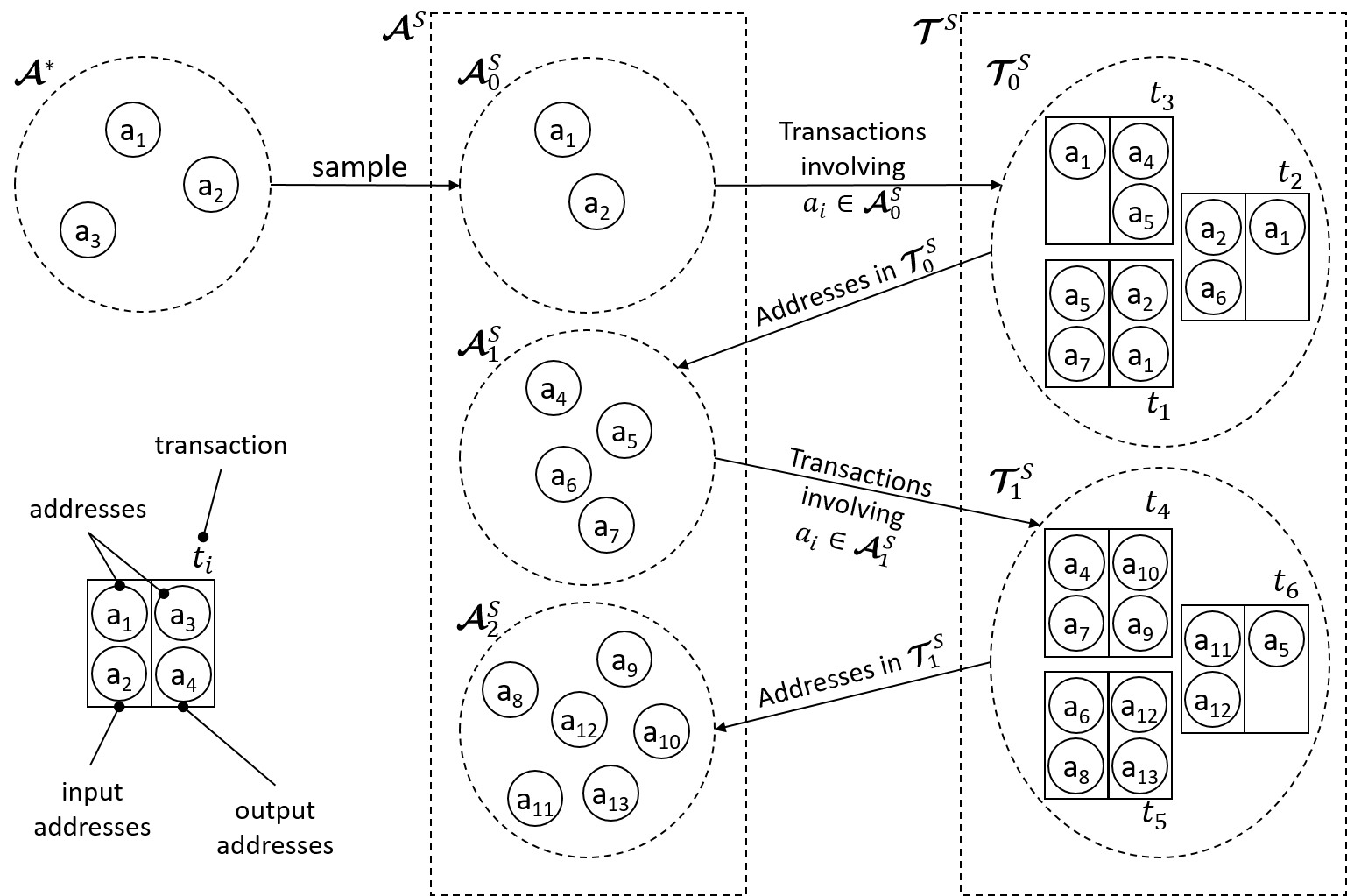}
    \caption{The transaction sampling process.}
    \label{fig:sampling}
\end{figure}

For computational feasibility, we restrict our analysis to a sample of \strans, as depicted in Figure \ref{fig:sampling}. 
First, we randomly select a subset $\ssamplei{0} \subseteq \sgt$ of the addresses in the ground truth. Next, we select all transactions involving an address $\addr \in \ssamplei{0}$ as an input or output, i.e., $\stransami{0} = \{\trans \:|\: \exists \addr \in \ssamplei{0} : \addr \in \mtrain{\trans} \cup \mtraout{\trans}\}$.
We then build the set \ssamplei{1} of addresses that appear in transactions of \stransi{0} but not in \ssamplei{0}, i.e. $\ssamplei{1} = \{\addr \:|\: \addr \not\in \ssamplei{0} \land \exists \trans \in \stransami{0} : \addr \in \mtrain{\trans} \cup \mtraout{\trans}\}$. 
The aforementioned process is then repeated in a similar manner. This involves finding the set \stransami{1} of transactions which include at least two addresses in \ssamplei{1}, i.e. $\stransami{1} = \{\trans \:|\: \trans \not\in \stransami{0} \land \exists\addri{1},\addri{2} \in \ssamplei{1} : \addri{1} \in \mtrain{\trans} \cup \mtraout{\trans} \land \addri{2} \in \mtrain{\trans} \cup \mtraout{\trans} \land \addri{1}\not=\addri{2}\}$. 
We set the condition on two addresses per transaction to reduce the size of the subsequently constructed Address Correspondence Networks.
Finally, we build \ssamplei{2} as the addresses appearing in transactions of \stransami{1} and not already in \ssamplei{0} or \ssamplei{1}, i.e. $\ssamplei{2} = \{\addr \:|\: \addr \not\in \ssamplei{0}\cup\ssamplei{1} \land \exists \trans \in \stransami{1} : \addr \in \mtrain{\trans} \cup \mtraout{\trans}\}$.

As a result, this process constructs a set of sampled transactions $\stransam = \stransami{0} \cup \stransami{1}$ having addresses $\ssample = \ssamplei{0}\cup \ssamplei{1} \cup \ssamplei{2}$. 
An advantage of this sampling method is that the constructed Address Correspondence Networks are centred around ground truth seed addresses, thereby exploiting the previous knowledge of controlling entities. 

\subsection{Partial and cumulative transaction sets} \label{s:method:subsets}
\begin{figure}[t!]
    \centering
    \includegraphics[width=0.75\textwidth]{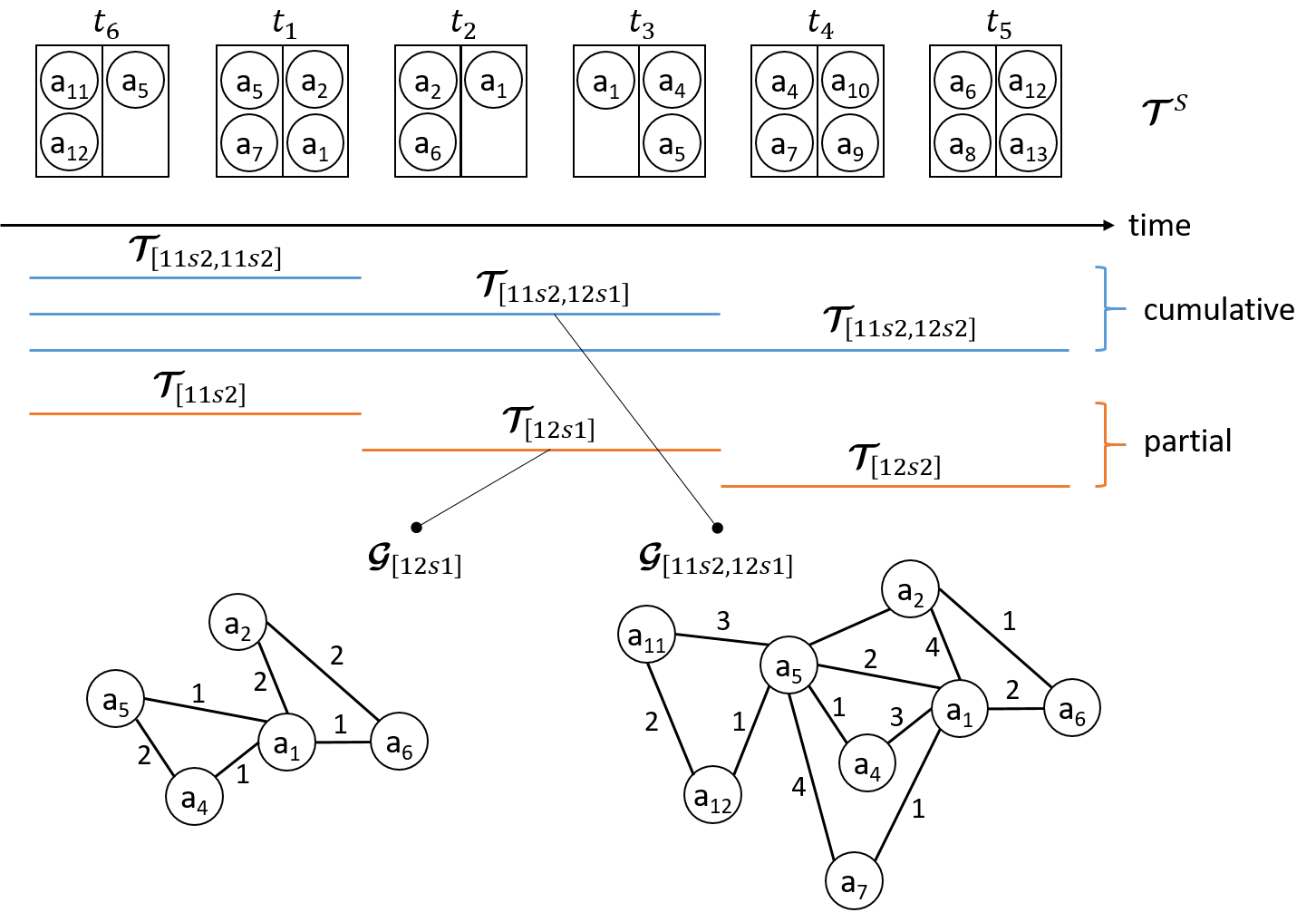}
    \caption{Cumulative and partial transaction sets, and construction of the Address Correspondence Networks.}
    \label{fig:networks}
\end{figure}

To study the evolution of the Bitcoin Address Correspondence Network over time, we create temporal subsets of the transactions in $\stransam$. Each subset includes only the transactions in \stransam that were generated in a specific time interval.
We create time intervals using two different strategies, which we name \emph{cumulative} and \emph{partial}, summarised in Figure \ref{fig:networks}.

The cumulative strategy creates eight time intervals of progressively increasing width\footnote{We represent dates in the use the DD.MM.YY format.}, $\{[01.07.11, 30.06.y], [01.07.11, 31.12.y] \:|\: y \in [12,15]\}$, while the partial strategy creates eight time intervals of fixed width, $\{[01.01.y, 30.06.y], [01.07.y, 31.12.y] \:|\: y \in [12,15]\}$.
It follows that cumulative time intervals overlap, while partial time intervals are disjoint.

Cumulative transaction sets are denoted with \stransamcum{y}{s}, which refers to all transactions in \stransam that were generated between the second semester of $2011$ and the $s^{\text{th}}$ semester of $y$, e.g. \stransamcum{14}{1} includes transactions generated in the interval $[01.07.11, 30.06.14]$.
Partial transaction sets are denoted with $\stransamocs{y}{s}{y}{s} \equiv \stransampar{y}{s}$, e.g. \stransampar{14}{1} refers to transactions generated in the interval $[01.01.14, 30.06.14]$. It is worth noting that while partial transaction sets do not share transactions, they may still share addresses which are used in multiple transactions.


\begin{figure}[t!]
    \centering
    \includegraphics[width=0.7\textwidth]{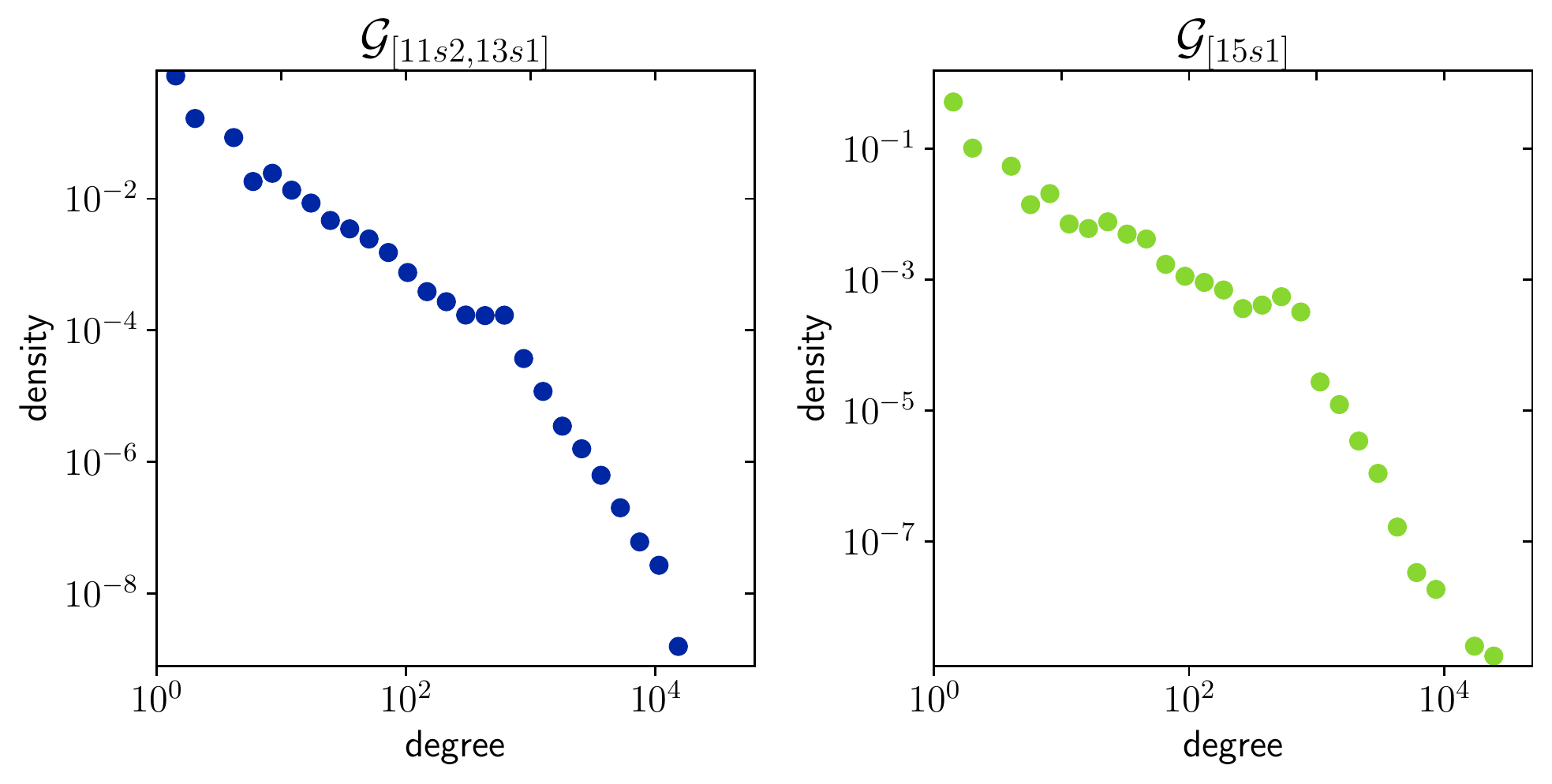}
    \caption{Degree distribution for cumulative \gcorneti{\intoc{11}{2}{13}{1}} and partial \gcorneti{\into{15}{1}}.}
    \label{fig:degree_distribution}
\end{figure}

\subsection{Address Correspondence Network construction} \label{s:method:corresp}

Let $\mwghn:\saddr\times\saddr\rightarrow\mathbb{N}$ be a function that counts how often an address pair, $(\addri{1}, \addri{2})$, is detected by any of the seven heuristics introduced in section \ref{s:bkg:heu} as being controlled by the same entity (considering only transactions in \stransamoc{o}{c}).
It is worth noting that \mwghn is symmetric (or undirected), i.e. $\mwgh{\addri{1}, \addri{2}} = \mwgh{\addri{2}, \addri{1}}$.

\begin{table}[b!]
\renewcommand{\arraystretch}{1.3}
\centering
\begin{tabular}{@{}ccccccccc@{}}
\toprule
\multicolumn{2}{c}{} & \multicolumn{3}{c}{Partial: $\gcorneti{\into{y}{s}}$} & & \multicolumn{3}{c}{Cumulative: \gcorneti{\intoc{11}{2}{y}{s}}}\\ 
\cmidrule{3-5} \cmidrule{7-9}
$y$& $s$ & $|\ssamplei{\into{y}{s}}|$  & $|\gcornetedgi{\into{y}{s}}|$      & $|\ssamplegti{\into{y}{s}}|$   & & $|\ssamplei{\intoc{11}{2}{y}{s}}|$    & $|\gcornetedgi{\intoc{11}{2}{y}{s}}|$       &$|\ssamplegti{\intoc{11}{2}{y}{s}}|$\\ 
\midrule
\multirow{2}{*}{2012} &1   &12      &46          &10    & &3,750     &164,408     &1,553\\ 
                      &2   &5,054   &1,239,850   &5,029 & &8,804     &1,404,258   &6,582\\ 
\addlinespace[3pt]
\multirow{2}{*}{2013} &1   &131,252 &3,183,594   &39,161&  &139,918   &4,587,813   &45,613\\ 
                      &2   &191,453 &45,965,678  &155,449& &329,240   &50,552,843  &199,614\\ 
\addlinespace[3pt]
\multirow{2}{*}{2014} &1   & 360,002 &81,891,103 &268,228& &607,098   &131,854,323 &396,548\\ 
                      &2   & 505,748 &31,121,336 &233,609& &1,092,560 &162,948,611 &621,919\\ 
\addlinespace[3pt]
\multirow{2}{*}{2015} &1   & 232,781 &16,836,377 &120,191& &1,270,261 &179,725,740 &734,185\\ 
                      &2   & 990,117 &52,732,659 &211,174& &2,184,445 &232,416,368 &935,599\\ 
\bottomrule                      
\end{tabular}
\caption{Number of nodes, edges and ground truth addresses of the partial and cumulative Address Correspondence Networks for each semester from 2012 to 2015.}
\label{table:network_desc}
\end{table}

The information captured by applying $\mwghn$ to each pair of addresses in \ssamplei{[o,c]} is collected in \emph{Address Correspondence Networks}, defined as undirected weighted graphs $\gcorneti{[o,c]}=(\ssamplei{[o,c]},\gcornetedgi{[o,c]},\mwghn)$. 
The construction process is depicted in Figure \ref{fig:networks}.
The addresses in \ssamplei{[o,c]} are the vertices of the graph, and \mwghn is the weight function.
$\gcornetedgi{[o,c]} \subseteq \ssamplei{[o,c]} \times \ssamplei{[o,c]}$ is the set of edges connecting address in two ways:
\begin{enumerate}
    \item pairs $(\addri{i},\addri{o})$ such that it exists a transaction $\trans\in\stransamoc{o}{c}$ having respectively \addri{i} and \addri{o} in its input and output address sets \saddrini{t} and \saddrouti{t}, and having $\mwgh{\addri{i}, \addri{o}}>0$.
    \item pairs $(\addri{i_1},\addri{i_2})$  such that it exists a transaction $\trans\in\stransamoc{o}{c}$ having both \addri{i_1} and \addri{i_2} in its input set \saddrini{t}, and having $\mwgh{\addri{i_1}, \addri{i_2}}>0$.
\end{enumerate}
%
Note that in a transaction, different heuristics can concur by identifying the same address as a change address, increasing the weights of the edges related to such an address. Figure \ref{fig:degree_distribution} shows the degree distribution of the Address Correspondence Networks \gcorneti{[11s2,12s1]} and \gcorneti{[11s2]}. The two distributions show a similar shape, but note that the left plot is a cumulative graph and the right plot is a partial graph; this indicates that the correspondence networks appear to preserve common properties across time. Table \ref{table:network_desc} provides descriptive statistics of the 16 Address Correspondence Networks we constructed from the eight partial and cumulative transaction sets. While the degree distributions cannot be assimilated to a single statistical distribution, they are skewed and fat-tailed, features that are recognised in complex networks of different contexts like biological, technological or social interactions \cite{barabasi_scale-free_2003}.

\begin{figure}[t!]
    \centering
    \includegraphics[width=1\textwidth]{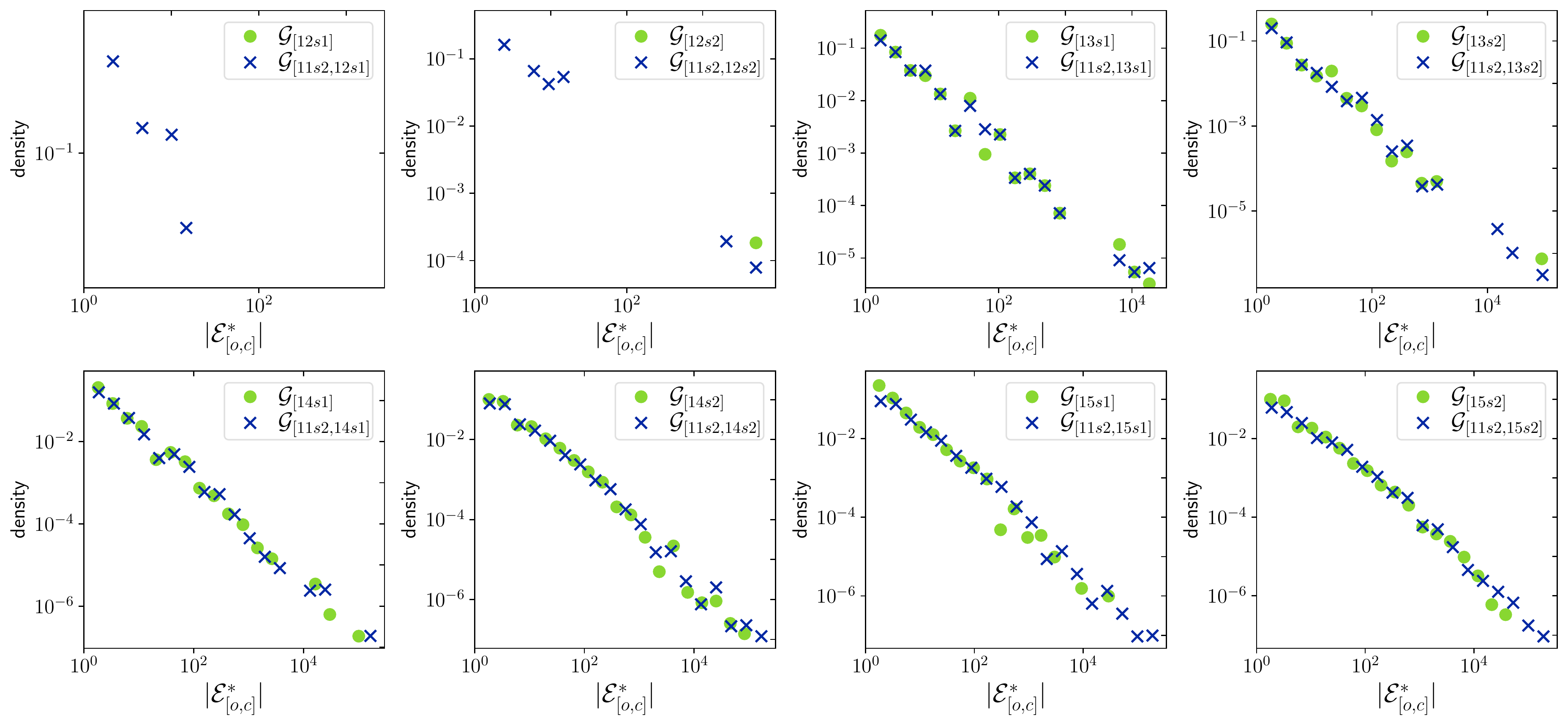}
    \caption{Distribution of ground truth entity sizes, $|\sentgti{[o,c]}|$.}
    \label{fig:entity_size_distribution}
\end{figure}

Figure \ref{fig:entity_size_distribution} shows the distribution of ground truth entities in the Address Correspondence Networks. In each plot, we compare a cumulative network and the partial network from its last six months, e.g. \gcorneti{[11s2,13s1]} with \gcorneti{[13s1]}. The number of known entities in the networks from 2012 is small, \gcorneti{[12s1]} and \gcorneti{[12s2]} do not show any relation with their pairs. However, from 2013, the similarity between distributions of known entities of partial and cumulative networks is notorious.

\subsection{Address Correspondence Network clustering} \label{s:method:cluster}

\begin{figure}[t!]
    \centering
    \includegraphics[width=0.75\textwidth]{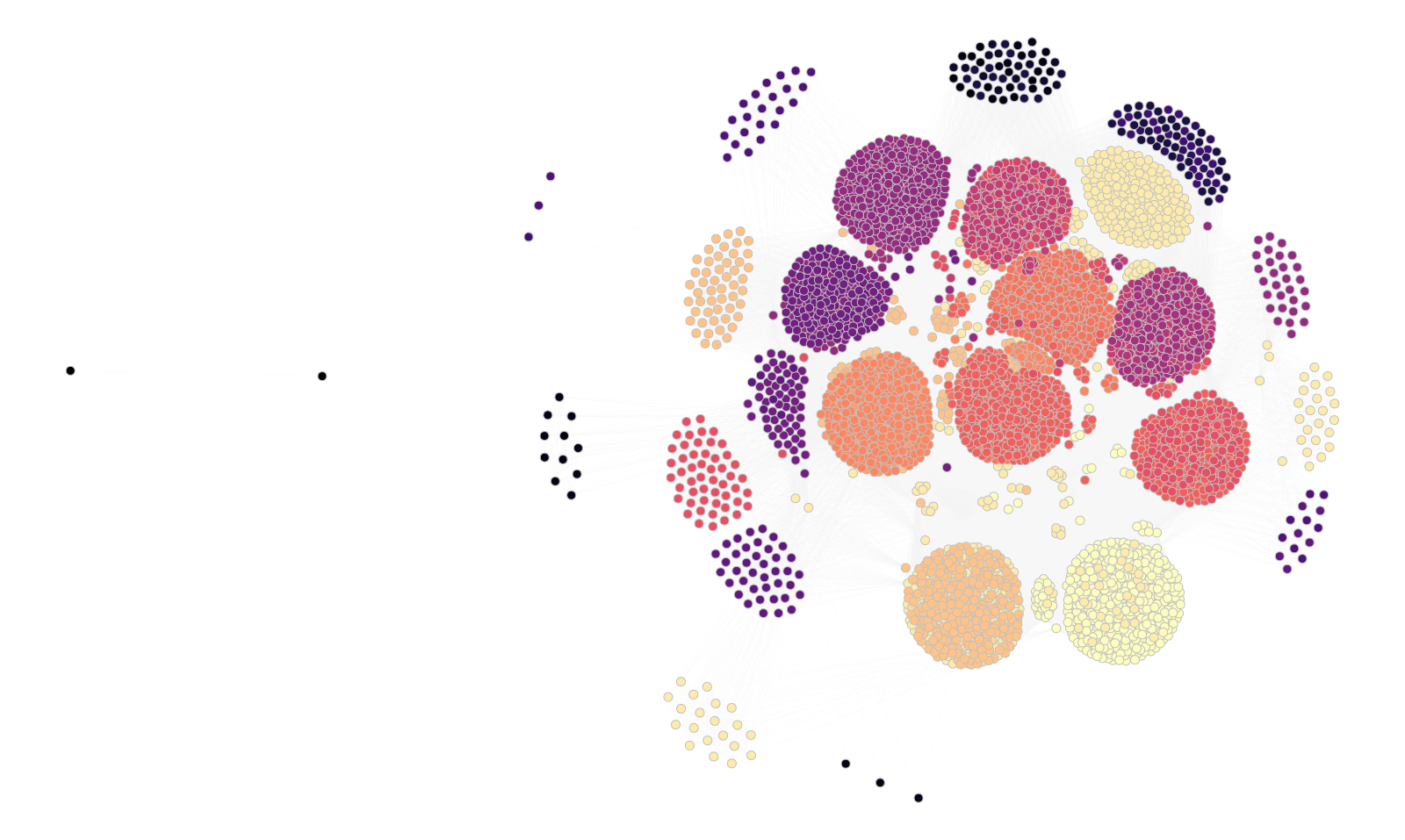}
    \caption{The Address Correspondence Network $\gcorneti{[12s2]}$. Clusters are identified by color.}
    \label{fig:example}
\end{figure}

Let $\gcorneti{[o,c]}= (\ssamplei{[o,c]},\gcornetedgi{[o,c]},\mwghn)$ be the Address Correspondence Network for the time interval $[o,c]$. We approach the entity identification problem by applying a community detection algorithm to \gcornetedgi{[o,c]} (therefore assuming that communities are sets of addresses belonging to the same entity). In $\gcorneti{[o,c]}$, highly interconnected vertices are clusters (communities) of addresses linked by one or several heuristics. Community detection algorithms find clusters of vertices highly interconnected but with sparse links between clusters. Specifically, the Label Propagation Algorithm (LPA) by \cite{raghavan_near_2007} finds communities and has linear complexity on the number of edges $\mathcal{O}(\gcornetedgi{[o,c]})$. The comparative study by \cite{yang_comparative_2016} shows that the scalability of LPA outperforms other fast clustering algorithms, including Leading Eigenvector by \cite{newman_finding_2006}, Walktrap by \cite{pons_computing_2006}, and Multilevel by \cite{blondel_fast_2008}. In LPA, each node is initialised with a unique label, denoting the cluster it is part of (the controlling entity of an address). In the basic case, all the nodes are initially assigned a random label.
Afterwards, each node is randomly visited and assigned a label according to the majority voting of its neighbours. The process repeats until every node in the network gets a label to which most of its neighbours belong.
Figure \ref{fig:example} shows a clustering for the partial network $\gcorneti{[12s2]}$. 

To initialise parts of the nodes, we use the information from the ground truth \maddentgtn. 
Let $\ssamplegti{[o,c]}$ denote the set of ground truth addresses in \gcorneti{[o,c]}, i.e. $\ssamplegti{[o,c]} = \ssamplei{[o,c]} \cap \sgt$, and let \sentgti{[o,c]} be the set of ground truth entities in \gcorneti{[o,c]}, i.e. $\sentgti{[o,c]} = \{\maddentgt{\addr} | \exists\addr\in\ssamplegti{[o,c]}\}$.
We assign to a subset of nodes $\ssampleinii{[o,c]}\subseteq\ssamplegti{[o,c]}$ the label from the ground truth, i.e. \maddentgt{\addr}.
%
It holds that \(\ssampleinii{[o,c]}  \subseteq  \ssamplegti{[o,c]} \subseteq \ssamplei{[o,c]}\)  and, concomitantly,   \( |\ssampleinii{[o,c]}| \leq |\ssamplegti{[o,c]}| \leq |\ssamplei{[o,c]}| \). 

In this paper, we are interested in exploring the ability of community detection algorithms to provide additional information about the true identities of users. We hypothesise that the Address Correspondence Network encodes additional information about the entities that control specific addresses. We argue that successive applications of heuristics may lead to connections between addresses controlled by the same entity that are denser and higher weighted than connections between addresses of different entities. 
Following this argument, we apply LPA to obtain a disjoint set of clusters $\sclus_{[o,c]} = \{\sclusiti{1},\ldots,\sclusiti{k}\}$,  such that $\bigcup_{i=1}^{k} \sclusiti{i} = \ssamplei{[o,c]}$. 
Because of the additional information provided by the ground truth, we modified LPA to avoid that the addresses in \ssampleinii{[o,c]} can change label, as they are associated with the actual entity according to the ground truth information. 

In the experiments, we vary the proportion $\msampleinii$ of initialised nodes, that is defined as:
\[ \msampleinii = {|\ssampleinii{[o,c]}|}/{|\ssamplei{[o,c]}|}.\]
Since \( {|\ssamplegti{[o,c]}|}/{|\ssamplei{[o,c]}|}\) varies across networks and is an upper bound on the proportion of initialised nodes, the domains of the approximated functions also vary.

\subsection{Cluster quality analysis} \label{s:method:quality}
Finally, we quantify the clustering quality as a function of cluster size and entity size. 
Given an Address Correspondence Network \gcorneti{[o,c]} and set of clusters $\sclus_{[o,c]} = \{\sclusiti{1},\ldots,\sclusiti{k}\}$ produced by LPA, we analyse the quality of $\sclus_{[o,c]}$ by defining a set of discrete random variables to describe characteristics of the network, and by five metrics: modularity to give information about the intrinsic quality of the clusters (and inherent topological structure of the network), homogeneity, entropy, Adjusted Mutual Information (AMI) and Adjusted Rand Index (ARI) to compare the clusters with the ground truth labels. Furthermore, all metrics are measured as functions of the proportion of initialised nodes $\msampleinii$.

\subsubsection{Random variables}\label{s:method:quality:vars}
To study the characteristics of the network, we define the following discrete random variables associated with the distributions of entities, addresses, and known addresses in the address correspondence network.

The first random variable, $\vE$, assumes a  value from the set of entities according to their frequency in the correspondence network. 
More specifically, $\vE$ can assume the value $\ent \in \sentgti{[o,c]}$ with probability equal to the numbers of addresses in \ssamplegti{[o,c]} mapped to \ent, divided by the total number of addresses in \ssamplegti{[o,c]}, i.e.:
$$
P(e) =\frac{|\{\addr\in\ssamplegti{[o,c]}|e=\maddentgt{\addr}\}|}{|\ssamplegti{[o,c]}|}.
$$

In addition to $\vE$, we also define variables that assume values in the entity set according to their frequency in specific clusters. 
Let $\vEi$ be the variable associated to the $i$-th cluster, i.e. $i\in[1,|\sclus_{[o,c]}|]$.
For each $i$, we build a histogram of the frequency of entities in \sclusiti{i}, by counting for each entity \ent the number of addresses associated to \ent through the ground truth data in \sclusiti{i}.
Such a histogram is used to approximate the distribution of entities over \sclusiti{i} and serves to describe $\vEi$.
Formally let $\ssampleigti{[o,c]}=\ssamplegti{[o,c]}\cap\sclusiti{i}$ be the set of addresses in \sclusiti{i} which are part of the ground truth.
$\vEi$ can assume a value \ent in $\sentigti{[o,c]}=\{\maddentgt{\addr}|\addr\in\ssampleigti{[o,c]}\}$ with probability:
$$
P(\ent) =\frac{|\{\addr\in\ssampleigti{[o,c]}|e=\maddentgt{\addr}\}|}{|\ssampleigti{[o,c]}|}.
$$

The variable $\vC$ assumes a cluster identifier according to its frequency over the addresses in the ground truth.
$\vC$ can assume a value $\sclusiti{i}\in\sclusi{[o,c]}$ with probability defined by the number of addresses in \ssamplegti{[o,c]} and \sclusiti{i} (i.e. \ssampleigti{[o,c]}) divided by the total number of addresses in \ssamplegti{[o,c]}, i.e.: 
\[P(\sclusiti{i}) = \frac{|\ssampleigti{[o,c]}|}{|\ssamplegti{[o,c]}|} .\]

Finally, we define variables complementary to \vEi to describe the frequency of clusters among each entity. 
We indicate with $\vCj$ the variable associated to the $j$-th entity \enti{j}, with $j\in[1,|\sentgti{[o,c]}|]$.
Given the entity \enti{j}, we build the histogram of the appearance of \enti{j} in each cluster of \sclusi{[o,c]}.
As for the $\vEi$ variables, we approximate the real distribution using the ground truth data, and considering only the addresses from $\sgt$ to build the bins. 
Formally, $C_j$ can assume values in $\sclusi{[o,c]}$ with probability: 
\[P(\sclusiti{i}) =
\frac{|\{\addr\in\sclusiti{i}|\enti{j}=\maddentgt{\addr}\}|}{|\{\addr\in\ssamplegti{[o,c]}|\enti{j}=\maddentgt{\addr}\}|}.\]

\subsubsection{Metrics}
\emph{Modularity}, initially proposed by \cite{newman_finding_2004}, compares the clusters with a random baseline.
This is done by computing the difference between the number of edges inside the clusters with the expected value of edges using the same clusters but with random connections between the nodes. 
Let $|\sclus_{[o,c]}|$ be the number of clusters in the Address Correspondence Network \gcorneti{[o,c]}, $q_{ij}$ the ratio of edges connecting addresses between cluster $\sclusiti{i}$ and cluster $\sclusiti{j}$, and $r_i=\sum_{j} q_{ij}$ the ratio of edges with at least one end in $\sclusiti{i}$. 
The modularity is defined as: 
\[Q = \sum_{i=1}^{|\sclus_{[o,c]}|} (q_{ii}-r_i^2).\]
A value close to 0 indicates that the community structure is akin to a random network, while values close to 1 indicate strong community structures, meaning dense connections inside the communities and sparse connections between them.

\emph{Information Theory Metrics.} Entropy, introduced in an information theory context by \cite{shannon_mathematical_1948}, quantifies the expected amount of information or uncertainty contained in a random variable.
Let $X$ be a discrete random variable, 
which can assume values $\{x_1, x_2, \ldots, x_k\}$ 
with probability $\{P(x_1), P(x_2), \ldots, P(x_k)\}$.
The entropy of $X$ is defined as: 
\[H(X) = -\sum_{x \in 1}^k P(x) \log_2 P(x),\]
while the normalised Shannon entropy is:
\[\hat{H}(X)=\frac{H(X)}{H_{max}(X)}=\frac{H(X)}{\log_2(k)}.\]
We use the normalised entropy of $E_i$ and $C_j$ to study the clusters by the perspective of the entities and the one of the cluster themselves.



Entropy also gives important information of the interrelation between random variables. 
Let us consider two variables $X$ and $Y$, 
and let $P(X,Y)$ be the joint probability distribution.
The \emph{conditional entropy} $H(Y|X)$ is defined as: 
\[H(Y|X) =  -\sum_{x \in X, y \in Y} P(x,y) \frac{\log_2 P(x,y)}{P(x)}\]
%
The conditional entropy indicates how much extra information is needed to describe $Y$ given that $X$ is known. 
Additionally, the amount of information needed on average to specify the value of two random variables is $H(X,Y) = H(X|Y)+H(Y)$.

We use conditional entropy to measure the quality of the clusters.
We do it by comparing them with the distribution of the entities in the Address Correspondence Network, exploiting the variables \vE and \vC. 
Such a measure is named \emph{homogeneity} and is initially introduced by \cite{rosenberg_v-measure_2007}.
Ideally, a cluster should only contain addresses that are controlled by the same entity. 
In such a case, clusters are homogeneous and it holds $H(\vE|\vC)=0$. 
The homogeneity score $h\in [0,1]$ is defined by: 
\[h = \begin{cases} 
      1 & \text{if} \: H(E,C)=0\\ 
      1 - {H(E|C)}/{H(E)}& \text{otherwise} \end{cases}.\]

The fundamental \emph{Mutual Information} (MI)  \cite{cover_elements_2006} quantifies the agreement between partitions. 
In addition to \sclusi{[o,c]}, let $\scls_{[o,c]} = \{\sclsiti{1},\ldots,\sclsiti{k}\}$ be an alternative set of clusters.
We introduce the variable $K$ to describe the distribution of the addresses in $\scls_{[o,c]}$, similarly to how we defined \vC for \sclusi{[o,c]} in Section \ref{s:method:quality:vars}.
The MI of \vC and $K$ is defined as: 
\[\hbox{MI}(\vC,K)=H(K)-H(K|\vC),\] 
and quantifies the reduction of the uncertainty of $\sclus_{[o,c]}$ due to the knowledge of $\scls_{[o,c]}$. 
The average MI value between $\sclus_{[o,c]}$ and $\scls_{[o,c]}$ tends to increase as the number of clusters increases, even if there is no difference in the clustering methodology, e.g. if the partitions are assigned clusters randomly. 
The \textit{Adjusted Mutual Information} defined by \cite{vinh_information_2010} takes into account the randomness using the expected value of MI $E[\hbox{MI}]$ and normalizes its value:
\[
\hbox{AMI}(\vC,K) = \frac{\hbox{MI}(\vC,K)-E[\hbox{MI}(\vC,K)]}{ \langle H(\vC,K) \rangle -E[\hbox{MI}(\vC,K)]}.\]
$\hbox{AMI}$ gets values in the $[0,1]$ interval, and when two partitions perfectly match, $\hbox{AMI}=1$.

Finally, we consider the \textit{Rand Index} (RI), initially proposed by \cite{rand_objective_1971}, which compares two set of clusters while ignoring permutations. 
Let \sclusi{[o,c]} and $\scls_{[o,c]}$ be two sets of clusters. Let $x(\sclusi{[o,c]},\scls_{[o,c]})$ be the number of pairs of addresses from the ground truth \ssamplegti{[o,c]} which are in the same cluster in \sclusi{[o,c]} and in the same cluster in $\scls_{[o,c]}$, i.e.: 
\begin{align*}
x(\sclusi{[o,c]},\scls_{[o,c]})=|\{(a_1,a_2)&|a_1,a_2\in \ssamplegti{[o,c]}, a_1 \not= a_2 \\
    &\land \exists \sclusiti{i} \in \sclusi{[o,c]} : a_1,a_2\in \sclusiti{i} \\
    &\land \exists \sclsiti{j} \in \scls_{[o,c]} : a_1,a_2\in \sclsiti{j}\}|,
\end{align*}
and let $y(\sclusi{[o,c]},\scls_{[o,c]})$ be the number of address pairs from the ground truth \ssamplegti{[o,c]} which are in different clusters of \sclusi{[o,c]} and in different clusters of $\scls_{[o,c]}$, i.e.: 
\begin{align*}
y(\sclusi{[o,c]},\scls_{[o,c]})=|\{(a_1,a_2) &|  a_1,a_2\in \ssamplegti{[o,c]}, a_1 \not= a_2 \\ 
& \land \exists \sclusiti{i},\sclusiti{j} \in \sclusi{[o,c]} : a_1 \in \sclusiti{i},a_2\in \sclusiti{j}, i\not=j \\
& \land \exists \sclsiti{k},\sclsiti{l} \in \scls_{[o,c]} : a_1\in \sclsiti{k},a_2\in \sclsiti{l}, k\not=l\}|,    
\end{align*}
The Rand Index is defined as:
\[\hbox{RI}(\sclusi{[o,c]},\scls_{[o,c]})=\frac{x(\sclusi{[o,c]},\scls_{[o,c]})+y(\sclusi{[o,c]},\scls_{[o,c]})}{|\ssamplegti{[o,c]}|\times(|\ssamplegti{[o,c]}|-1)},\] 
where the denominator is the number of address pairs in $\ssamplegti{[o,c]}$. 
As with MI/AMI, we consider an adjusted version of RI, the \textit{Adjusted Rand Index} (ARI) as proposed by \cite{hubert_comparing_1985}, which accounts for chance:
\[
\hbox{ARI}(\sclusi{[o,c]},\scls_{[o,c]}) = \frac{\hbox{RI}(\sclusi{[o,c]},\scls_{[o,c]})-E[\hbox{RI}(\sclusi{[o,c]},\scls_{[o,c]})]}{ \max\langle\hbox{RI}(\sclusi{[o,c]},\scls_{[o,c]})\rangle -E[\hbox{RI}(\sclusi{[o,c]},\scls_{[o,c]})]},\]

where $E[\hbox{RI}(\sclusi{[o,c]},\scls_{[o,c]})]$ denotes the expected value of $\hbox{RI}(\sclusi{[o,c]},\scls_{[o,c]})$. \cite{warrens_equivalence_2008} shows that ARI is equivalent to Cohen's Kappa (\cite{cohen_coefficient_1960}), which is well suited for the evaluation of community detection methods, as discussed by \cite{liu_evaluation_2020}.


\section{Results} \label{s:result}

\begin{figure}[b]
\centering
\includegraphics[width=0.7\textwidth]{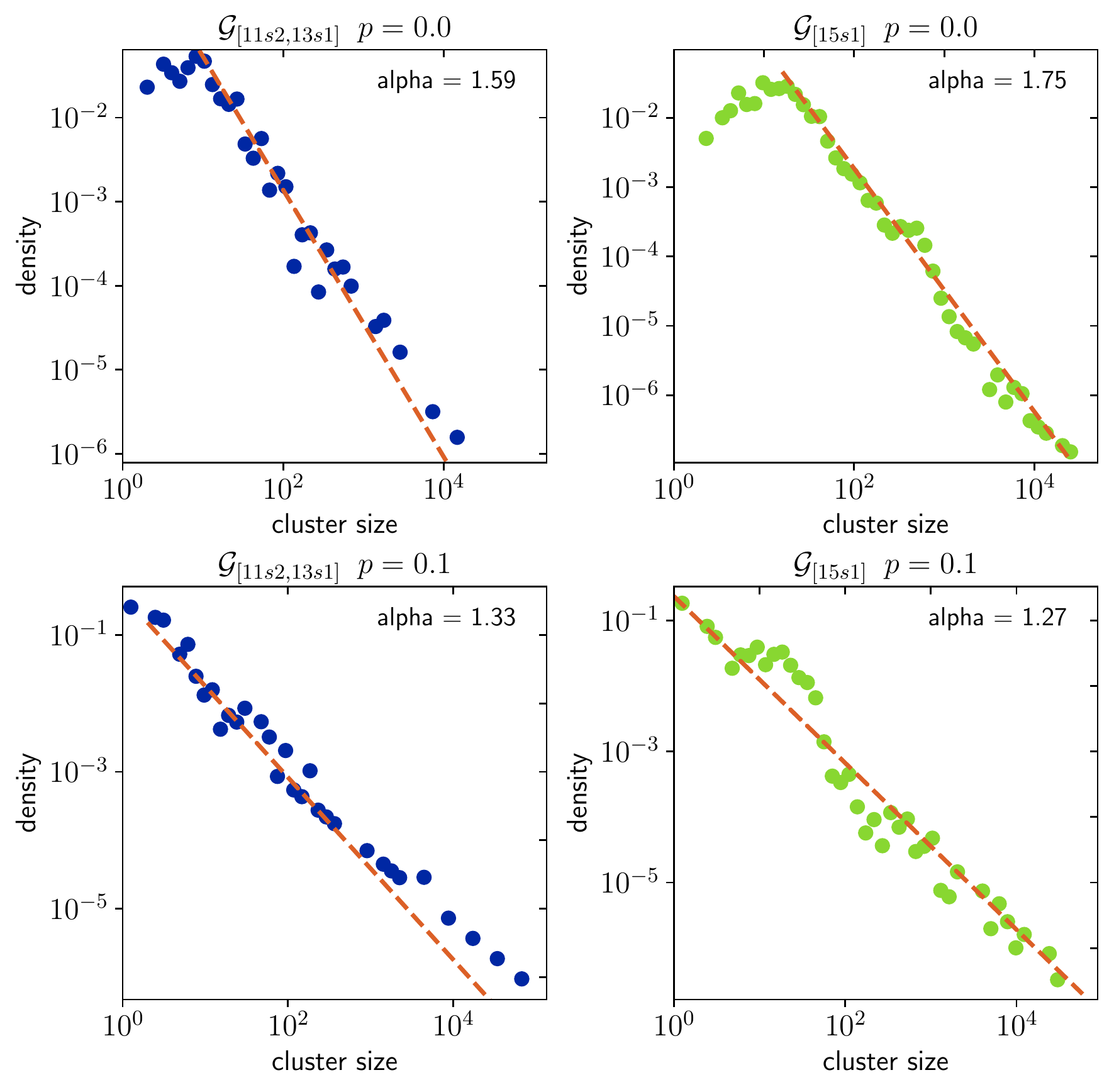}
\caption{Cluster size distribution of \gcorneti{[11s2,13s1]} and \gcorneti{[15s1]} for $\msampleinii=0.0$ and $\msampleinii=0.1$. The alpha values of the power-law distribution fits are also shown.}
\label{fig:cluster_size_distribution}
\end{figure}

We first analyse the size of the clusters identified by LPA for the Address Correspondence Networks described in Section \ref{s:method}, whose statistics are shown in Table \ref{table:network_desc}. Figure \ref{fig:cluster_size_distribution} shows the cluster size distribution of \gcorneti{[11s2,13s1]} and \gcorneti{[15s1]}, for initialisation proportions $\msampleinii=0$ and $\msampleinii=0.1$. Note that the density of the small clusters, in both cases, shifts to reach larger cluster sizes when $\msampleinii=0.1$, as well as the maximum cluster size of \gcorneti{[11s2,13s1]}. This indicates that even a small proportion of initialised nodes, such as $\msampleinii=0.1$,  considerably modifies the cluster distribution in the networks.

We also fit a power-law distribution to the cluster size distribution, shown by the dotted red lines with the corresponding alpha values in Figure \ref{fig:cluster_size_distribution}. Furthermore, the power-law distribution fits the data significantly better than an exponential distribution, resulting in $p$-values of less than $0.1\%$ using likelihood ratio tests \cite{clauset_power-law_2009}. The exponents are larger for $p=0$ than for $p=0.1$, in agreement with the observation related to the range of values in the cluster size. In general, the distributions are very heterogeneous. Additionally, the cluster size distribution suggests that, from a Correspondence Network perspective,  there is a preferential attachment dynamic in the address generation where entities that control many addresses are likely to generate more addresses than others.

\begin{figure}[t]
\centering
\includegraphics[width=0.7\textwidth]{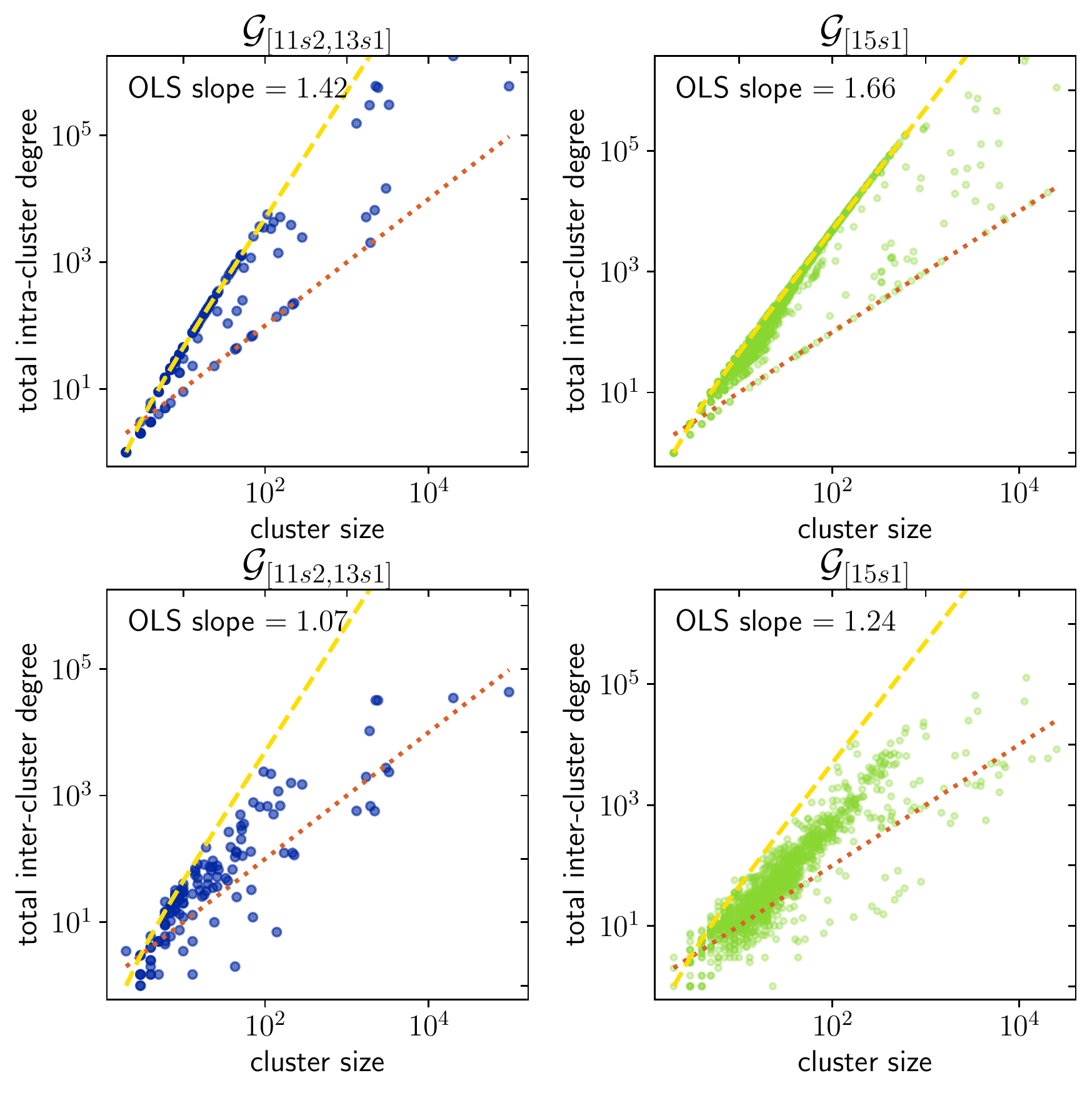}
\caption{Comparison of the total intra-cluster and inter-cluster degrees for \gcorneti{\intoc{11}{2}{13}{1}} and \gcorneti{\into{15}{1}}. We also show the lines $y=x$ (red, dotted) and $y={x(x-1)}/{2}$ (yellow, dashed).}
\label{fig:inter_intra_cluster_links}
\end{figure}

Next, we study the behaviour of the intra-cluster total degree (number of edges connecting nodes that belong to the same cluster) and the inter-cluster degree (number of edges between nodes that belong to different clusters) as functions of the cluster size. For the total intra-cluster degree, there are two extreme behaviours that can be expected. On the one hand, a linear dependency on cluster size would signal that address reuse is negligible (therefore that privacy-preserving usage are commonplace), and the topology of the correspondence network encodes no additional information about the identity of the users that control the addresses. On the other hand, a quadratic relationship (close to the theoretical maximum $\propto c (c-1)/2$) would signal that the clusters are very densely interconnected, and the actual address reuse is high. Therefore, it would be possible to infer actual information about the users by directly inspecting the correspondence network through network science methods.
In Figure \ref{fig:inter_intra_cluster_links}, the extreme values of the intra-cluster degree of \gcorneti{11s2,13s1} and \gcorneti{15s1} are above a linear function (red dotted line) and below a quadratic function (yellow dashed line) of the cluster size. The same lines are depicted in the inter-cluster degree distributions showing that the intra-cluster degree grows faster. By applying an Ordinary Least Squares regression (OLS), the slope of a fitting line is in both networks bigger in the intra-cluster case. Furthermore, bigger entities preserve this behaviour, showing that the correspondence network has an inherent community structure. Thus, this result is not valid only for entities that control a small number of addresses, and it follows that it is a general property of the network. 

\begin{figure}[t]
\centering
\includegraphics[width=1\textwidth]{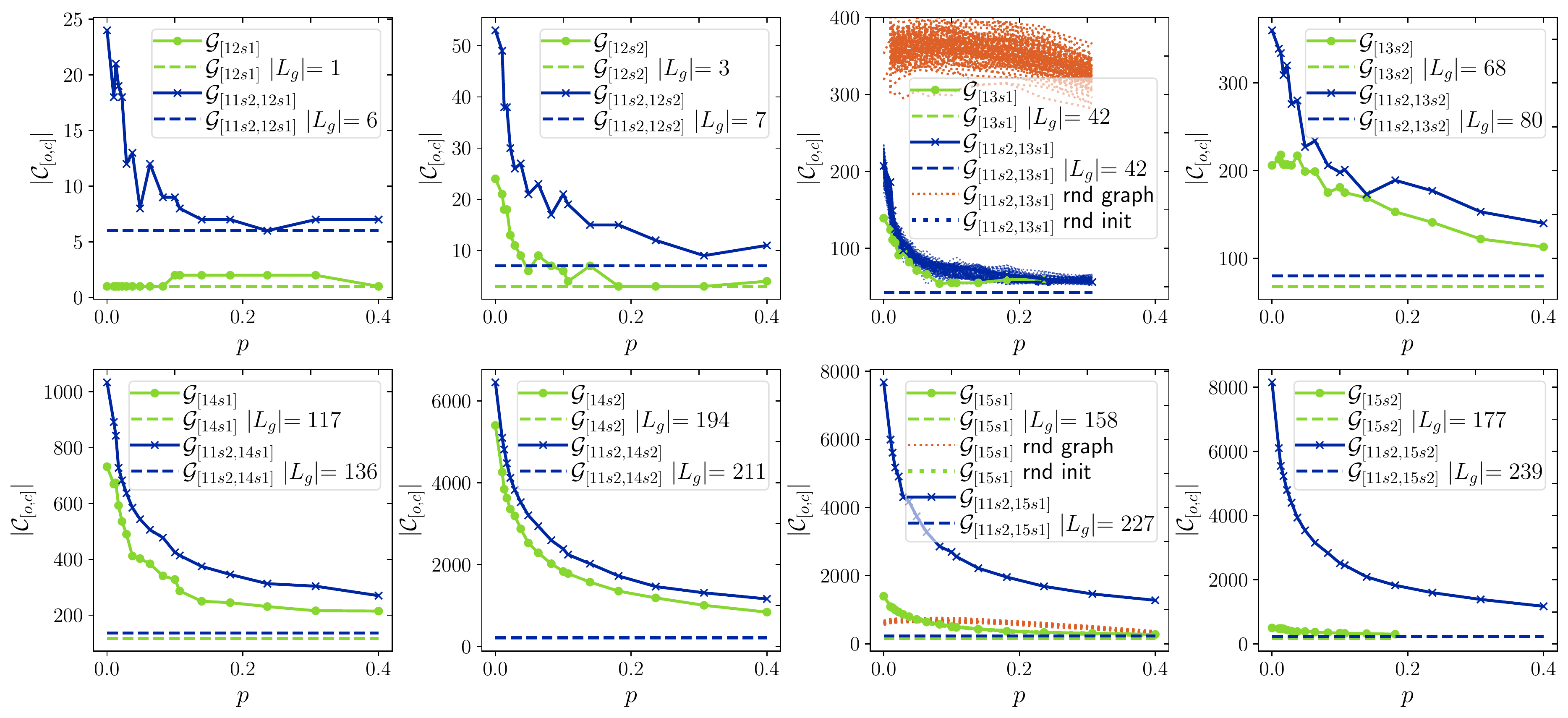}
\caption{Number of clusters as a function of $\msampleinii$.}
\label{fig:num_clusters}
\end{figure}

Figure \ref{fig:num_clusters} shows the number of clusters returned by LPA, $|\sclus_{[o,c]}|$, as a function of $\msampleinii$. 
The dashed lines indicate the number of entities $|\sentgti{[o,c]}|$ for each Address Correspondence Network. 
$|\sentgti{[o,c]}|$ is a lower bound of the true number of entities, since each network also contains addresses not in the ground truth. 
This is supported by $|\sclus_{[o,c]}| \geq |\sentgti{[o,c]}|$ holding for each test point. 
In general, $|\sclus_{[o,c]}|$ decreases sharply at small $\msampleinii$, after which the rate of decrease slows and stabilises. 
$|\sclus_{[o,c]}|$ tends to be lower for partial networks than for cumulative networks, and can be explained by partial networks having a lower $|\sentgti{[o,c]}|$.

\begin{figure}[h!]
\centering
\includegraphics[width=1\textwidth]{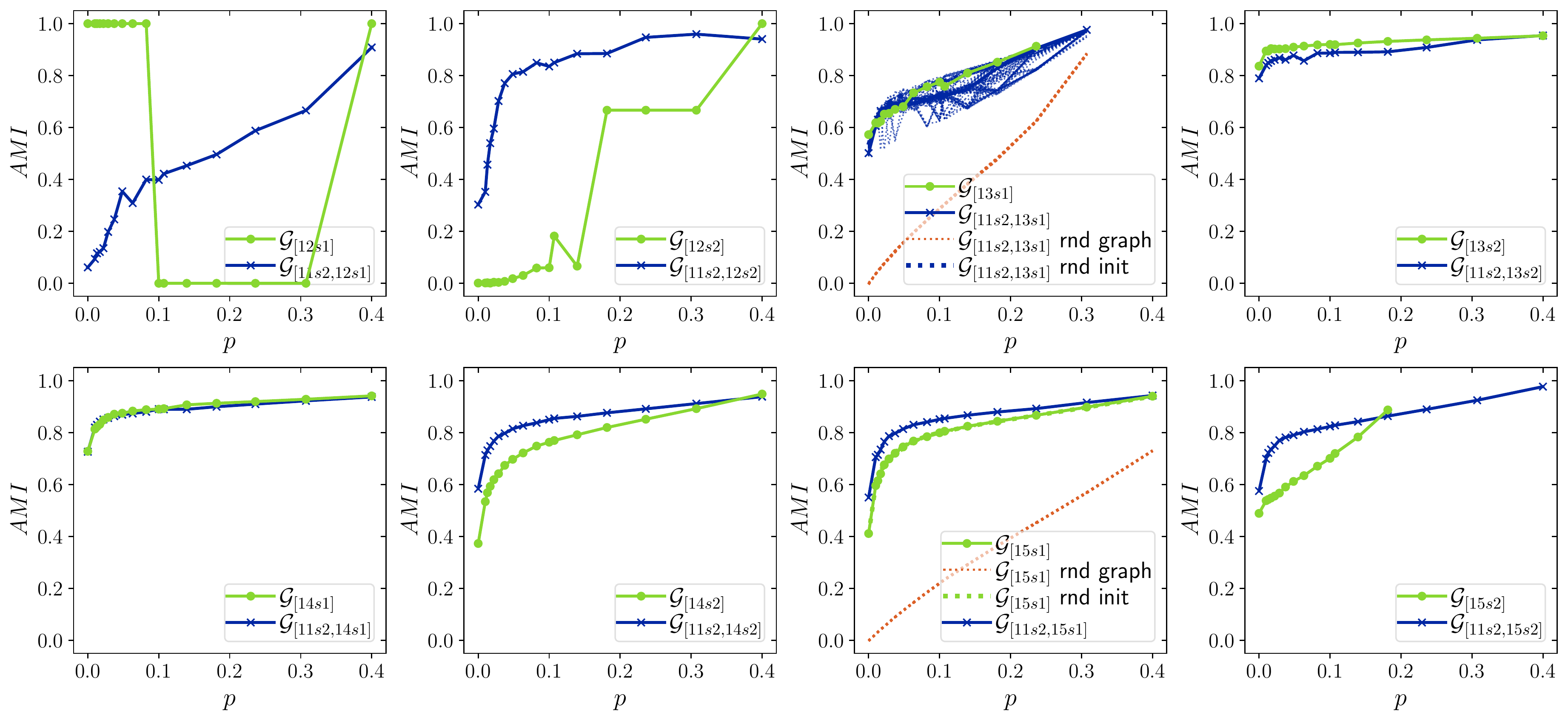}
\caption{AMI as a function of $\msampleinii$.}
\label{fig:adjusted_mutual_information}
\end{figure}

\begin{figure}[h!]
\centering
\includegraphics[width=1\textwidth]{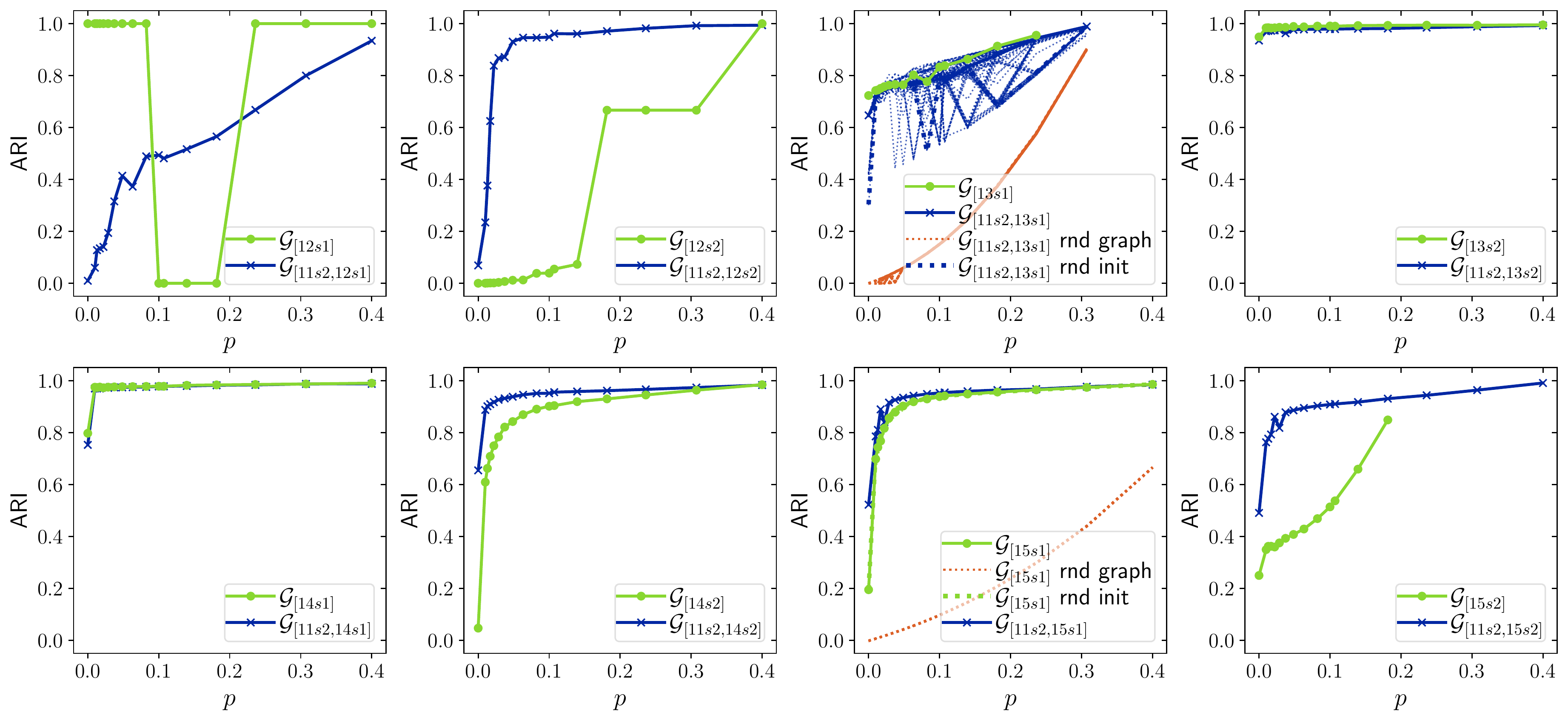}
\caption{ARI as a function of $\msampleinii$.}
\label{fig:adjusted_rand_index}
\end{figure}

\begin{figure}[h!]
\centering
\includegraphics[width=1\textwidth]{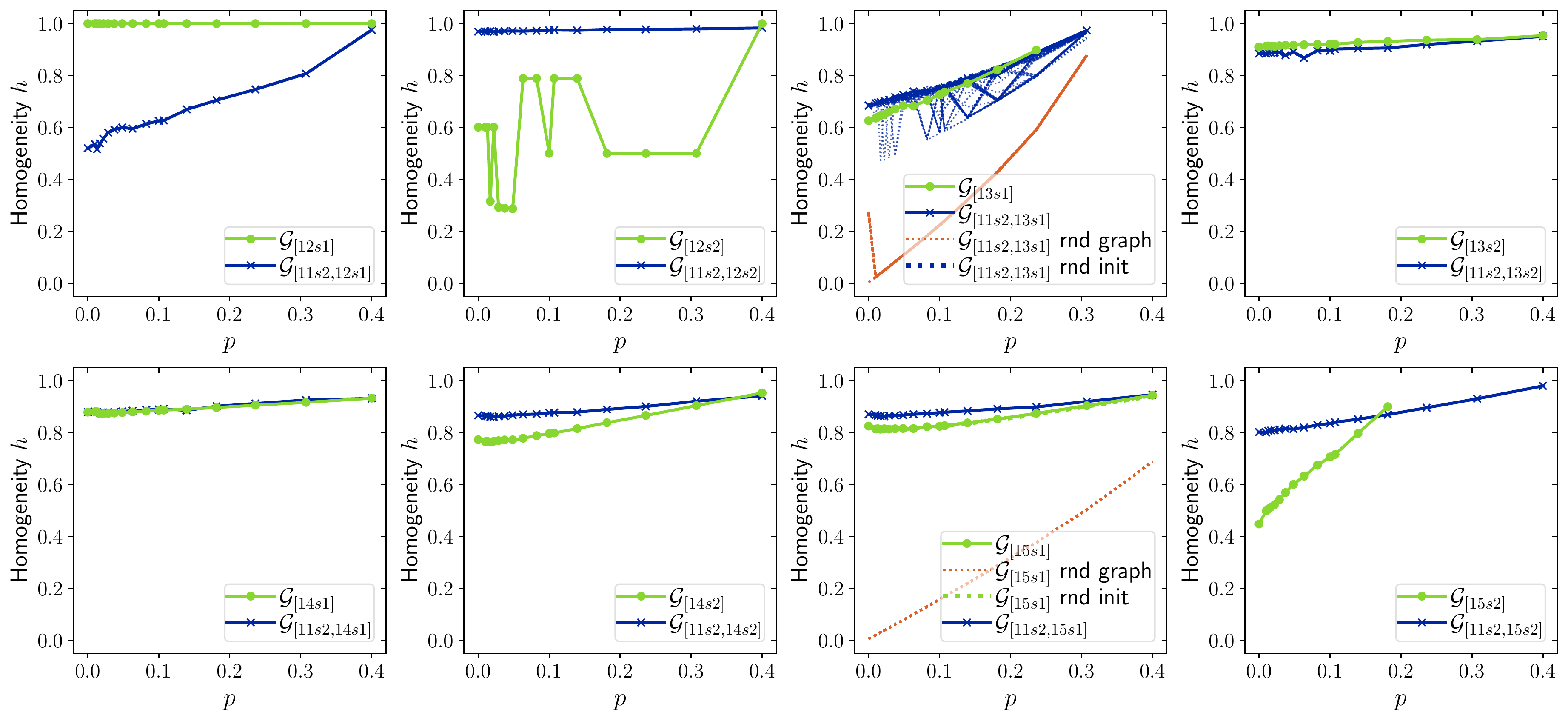}
\caption{Homogeneity as a function of $\msampleinii$.}
\label{fig:homogeneity}
\end{figure}

\emph{The complexity and structure of the Address Correspondence Network are stable over time.} Figures \ref{fig:adjusted_mutual_information}, \ref{fig:adjusted_rand_index} and \ref{fig:homogeneity} show AMI, ARI and homogeneity as functions of $\msampleinii$. Since these metrics require ground truth labels, they are computed only for addresses in $\ssamplegti{[o,c]}$. We observe that AMI and ARI lead to similar results: they rapidly increase before converging to the maximum value as $\msampleinii$ increases. In contrast, homogeneity exhibits no such initial rapid increase, and instead increases linearly with $\msampleinii$. The mean levels of AMI, ARI and homogeneity do not consistently increase or decrease with increasing half-year. Furthermore, the mean metric levels for the partial networks appear to be comparable to those for the cumulative networks. This suggests that the complexity and structure of the Address Correspondence Network communities remain stable over time. 

\emph{The effect of the node initialisation.} If the cost of labelling a Bitcoin address is assumed to be constant, the marginal gain in clustering quality per unit cost from increasing $\msampleinii$ quickly declines. Considering that homogeneity remains constant across all $\msampleinii$, it appears that increasing $\msampleinii$ is cost-effective until around $\msampleinii = 0.1$. At this point, \ssampleinii{[o,c]} contains most of the information required to describe the community structure. The observed saturations in $|\sclus_{[o,c]}|$, AMI and ARI suggest that increasing $\msampleinii$ beyond 0.1 adds only idiosyncratic community information, yielding little improvement in clustering quality. This is further confirmed by studying clustering modularity as a function of $\msampleinii$ in Figure \ref{fig:homogeneity}. Modularity appears mostly constant except for a sharp initial change, showing a robust community topology that is consistently detected after initialising a small proportion of nodes.

\begin{figure}[h!]
\centering
\includegraphics[width=1\textwidth]{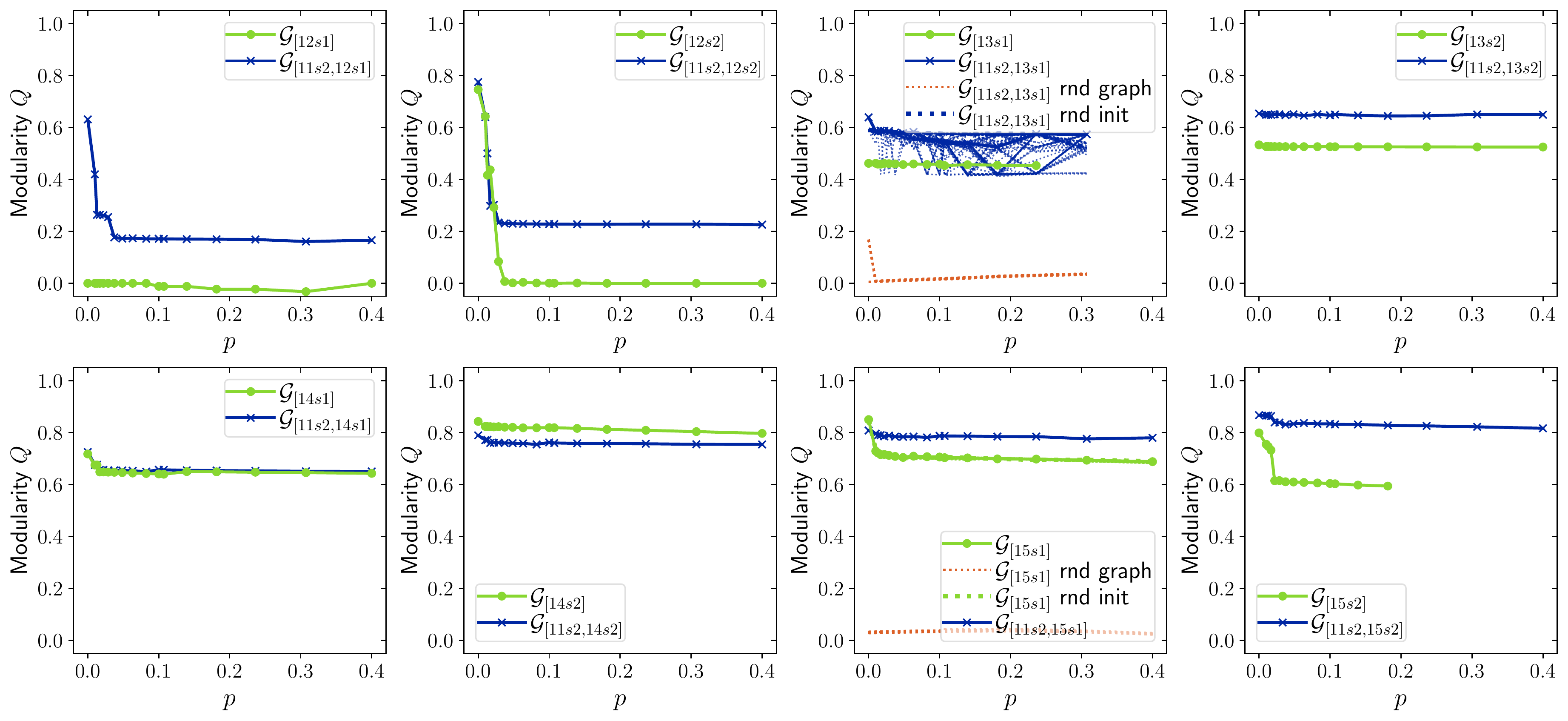}
\caption{Modularity as a function of $\msampleinii$.}
\label{fig:modularity}
\end{figure}

To assert the significance of the results presented in Figures \ref{fig:num_clusters} – \ref{fig:modularity}, we repeated the experiments for 100 randomised versions of the $\gcorneti{\intoc{11}{2}{13}{1}}$ and $\gcorneti{\into{15}{1}}$ Address Correspondence Networks. 
The $i-$th randomised network was obtained by performing $4i \cdot |\gcornetedgi{[o,c]}|$ edge swaps on the original network, according to the algorithm proposed by \cite{maslov_specificity_2002}, which preserves the network’s degree distribution. 
With the exception of $|\sclus_{\intoc{11}{2}{13}{1}}|$ for $\gcorneti{\intoc{11}{2}{13}{1}}$, the randomised results show little variation. However, all randomised results appear significantly different to those for the original networks. This suggests that the (non-randomised) results shown in Figures \ref{fig:num_clusters} – \ref{fig:modularity} 
are a consequence of more complex network properties rather than solely the degree distribution.

Furthermore, the effect of node initialisation order was studied by repeating the experiments for the $\gcorneti{\intoc{11}{2}{13}{1}}$ and $\gcorneti{\into{15}{1}}$ networks using 100 random orderings. The node initialisation order does not seem to affect the general level and shape of the curves. Small perturbations observed in Figures \ref{fig:num_clusters} – \ref{fig:modularity} appear to be idiosyncrasies of the chosen ordering, and may be larger for smaller networks (since the curves for $\gcorneti{\intoc{11}{2}{13}{1}}$ vary more than the ones for $\gcorneti{\into{15}{1}}$). 

\begin{figure}[t]
\centering
\includegraphics[width=1\textwidth]{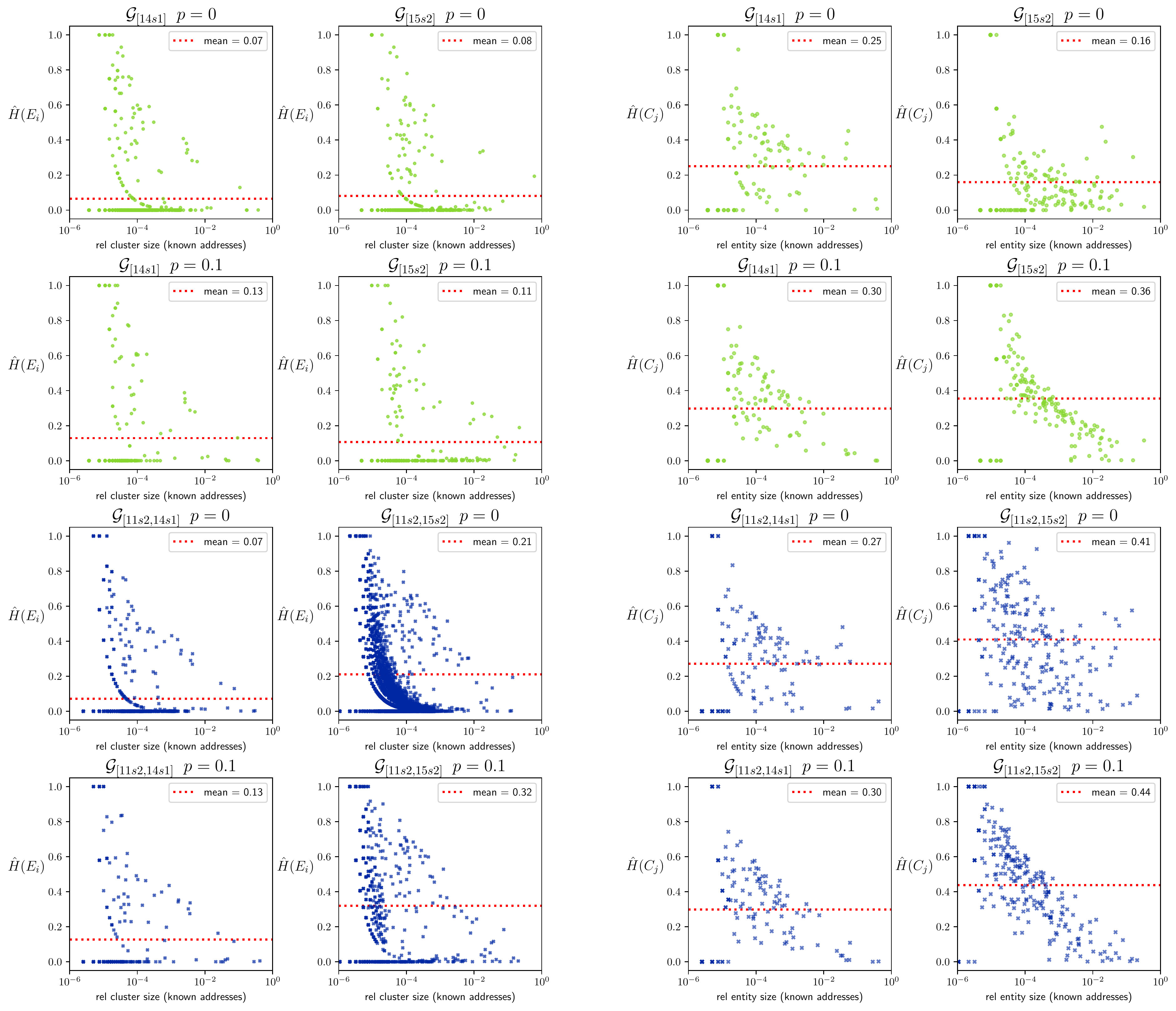}
\caption{Normalised entropy as a function of relative cluster size and relative entity size.}
\label{fig:entropies}
\end{figure}

\emph{The effect of cluster and entity sizes.}
Figure \ref{fig:entropies} shows $\hat{H}(E_i)$ and $\hat{H}(C_j)$ for the $\gcorneti{\into{14}{1}}$, $\gcorneti{\intoc{11}{2}{14}{1}}$, $\gcorneti{\into{15}{2}}$ and $\gcorneti{\intoc{11}{2}{15}{2}}$ networks.
$\hat{H}(E_i)$ and $\hat{H}(C_j)$ are expressed as functions of the relative cluster and entity sizes, i.e. normalised to  $|\ssamplegti{[o,c]}|$, respectively. 
We run experiments with $p=0$ and $p=0.1$. 
We note that \emph{$\hat{H}(E_i)$ correlates negatively with the relative cluster size, and $\hat{H}(C_j)$ correlates negatively with relative entity size}. For small clusters and entities, there are strips of points located at the minimum and maximum values of $\hat{H}(E_i)$ and $\hat{H}(C_j)$. 
This is to be expected: if we consider a cluster with only two addresses, both associated with the same entity, $\hat{H}(E_i)$ is minimum. 
If two addresses are mapped to different entities, we obtain a uniform entity label distribution, and $\hat{H}(E_i)$ is maximum. 
Such extreme fluctuations become less likely as cluster size increases. 
Large clusters, therefore, tend to be purer than smaller clusters, corresponding to a higher clustering quality. 
Similarly, entities represented by more addresses are distributed more asymmetrically across clusters, again corresponding to a higher clustering quality. 
This is in agreement with the results in Figure \ref{fig:inter_intra_cluster_links}, where the community structure is shown to become more apparent for larger clusters.

Furthermore, the mean levels of $\hat{H}(E_i)$ and $\hat{H}(C_j)$ for the partial networks are always less than or equal to the ones of the corresponding cumulative networks (comparing row 1 to row 3 and row 2 to row 4 in figure \ref{fig:entropies}). This suggests that partial networks allow a higher quality of interpretation regarding the community structure. 
A possible explanation for this is that Bitcoin entities have less time to obfuscate their activity: the longer the considered transaction history, the more the obfuscation attempts accumulate and the more difficult it becomes to detect the true community structure. 

Interestingly, the average $\hat{H}(E_i)$ and $\hat{H}(C_j)$ increase after initialising $10\%$ of nodes. 
The increase in $\hat{H}(E_i)$ can be explained by the loss of small, homogeneous clusters with low $\hat{H}(E_i)$. 
For $\hat{H}(C_j)$, the increase is likely due to the decrease in the number of clusters, which in turn causes $H_{max}(E_i)$ to decrease.


\section{Conclusion and future work} \label{s:conclusion}

In this paper, we consider the application of a general-purpose community detection algorithm, LPA, to detect address clusters that are controlled by the same entity in the Bitcoin transaction history. 
Specifically, we apply LPA to Address Correspondence Networks, which incorporate information from a variety of simple address linking heuristics. 
We detect a strong community structure within these networks by inspecting their intra- and inter-cluster degrees.
We find that the inter-cluster degree grows faster than the inter-cluster degree for cluster size increments. 
Address correspondence networks are therefore suitable for the application of general community detection methods from the broader field of network science---this creates an entry point for future researchers to move far beyond the application of primitive heuristics. 

Since LPA is able to exploit ground truth information, we find that clustering quality improves as the number of labelled addresses in the Address Correspondence Networks increases. 
However, under the assumption that the cost of labelling a Bitcoin address is constant, we find that the marginal gain in clustering quality per unit cost quickly declines. Under this assumption, we propose that address labelling is cost-effective until around $p=0.1$, i.e. until 10\% of all addresses in the Address Correspondence Network are identified. 
Furthermore, we find that choosing which addresses to label does not have a significant effect on clustering quality. 
Finally, we find that the structure of communities in the Address Correspondence Network remains stable over time. Partial Address Correspondence Networks are, therefore,  reasonable proxies for their cumulative counterparts (and far less demanding from a computational point of view).

For future work, we plan to conduct experiments to test the robustness of the heuristics and specific combinations between them.  For example, analysing their likelihood and studying their contribution to the links between addresses. From a network reconstruction perspective, link prediction is an interesting approach to improve the correspondence network by validating current links and predicting missing ones. Additionally, different machine learning approaches can be implemented to graph analysis; supervised methods are suitable if more ground truth information is available in the future.


\section*{Conflict of Interest Statement}

The authors declare that the research was conducted in the absence of any commercial or financial relationships that could be construed as a potential conflict of interest.

\section*{Author Contributions}
DDA and CJT conceived the experiment. JAF and AP performed the analysis. All authors discussed the methods and results. JAF and AP wrote a first draft. All authors worked and agreed on the final version.

\section*{Funding}
DDA acknowledges partial funding from by the Swiss National Science foundation under contract \# 407550\_167177. CJT acknowledges financial support from the  University of Zurich through the University Research Priority Program on Social Networks. 
 


\section*{Data Availability Statement}
The data analysed in this study is publicly available by synchronising the Bitcoin blockchain. 
The ground truth dataset is available at \href{https://www.walletexplorer.com/}{https://www.walletexplorer.com/}.

\bibliography{references.bib}

\end{document}